\documentclass[%
reprint,
superscriptaddress,
 amsmath,amssymb,
 aps,
 prb,
]{revtex4-2}

\usepackage{graphicx}
\usepackage{dcolumn}
\usepackage{bm}
\usepackage[normalem]{ulem}
\usepackage{xcolor}
\usepackage{mathrsfs}
\usepackage[T1]{fontenc}

\begin{document}

\title{Phase noise analysis and control of VO$_2$-based relaxation type oscillators}

\author{Artem Litvinenko} \thanks{Correspondence to: litvinenko@oakland.edu}
\affiliation{Department of Physics, University of Gothenburg, Gothenburg 41296, Sweden}
\affiliation{Department of Physics, Oakland University, Rochester 48309, USA}

\author{Erbin Qiu} \thanks{Correspondence to: e3qiu@ucsd.edu}
\affiliation{UCSD, USA}

\author{Sambit Ghosh}
\affiliation{Department of Physics, University of Gothenburg, Gothenburg 41296, Sweden}

\author{Juan Andres Hofer}
\affiliation{UCSD, USA}

\author{Akash Kumar}
\affiliation{Department of Physics, University of Gothenburg, Gothenburg 41296, Sweden}
\affiliation{Center for Science and Innovation in Spintronics and Research Institute of Electrical Communication (CSIS), Tohoku University, 2-1-1 Katahira, Aoba-ku, Sendai 980-8577 Japan}
\affiliation{Research Institute of Electrical Communication (RIEC), Tohoku University, 2-1-1 Katahira, Aoba-ku, Sendai 980-8577 Japan}

\author{Jong-Guk Choi}
\affiliation{Department of Physics, University of Gothenburg, Gothenburg 41296, Sweden}

\author{Ivan K. Schuller}
\affiliation{UCSD, USA}

\author{Johan \r{A}kerman}
\affiliation{Department of Physics, University of Gothenburg, Gothenburg 41296, Sweden}
\affiliation{Center for Science and Innovation in Spintronics and Research Institute of Electrical Communication (CSIS), Tohoku University, 2-1-1 Katahira, Aoba-ku, Sendai 980-8577 Japan}
\affiliation{Research Institute of Electrical Communication (RIEC), Tohoku University, 2-1-1 Katahira, Aoba-ku, Sendai 980-8577 Japan}

\date{\today}

\begin{abstract}
VO$_2$-based relaxation oscillators form a rapidly developing field that finds applications in neuromorphic computing, Ising machines, and numerous signal processing concepts. These oscillators operate in a deeply nonlinear relaxation regime based on rapid phase transitions between insulating and metallic states in the VO$_2$ material. This process is governed by thermal effects, which lead to additional voltage fluctuations and contribute to a considerably wide spectral linewidth in the VO$_2$-based oscillator signal. In this work, we thoroughly study the phase noise in VO$_2$-based relaxation oscillators and demonstrate that the broadening of the generation spectrum linewidth at low oscillation frequencies is caused by an increased susceptibility to thermal fluctuations during the incubation phase. We explore the types of noise affecting oscillator stability and show that synchronization with an external square-wave signal improves the phase noise more effectively than a sinusoidal-shape injection locking signal.

\begin{description}
\item[PACS numbers: 85.75.-d, 05.45.Xt, 75.40.Gb, 75.47.-m, 84.30.Qi]

\end{description}

\end{abstract}

\maketitle

\section{Introduction}
VO$_2$-based oscillators have emerged as highly reproducible and CMOS-compatible nanoscale oscillators exploited in a wide range of applications, ranging from signal processing and neuromorphic computing to brain-inspired artificial intelligence and unconventional computing systems, including oscillator-based Ising machines \cite{maher2024cmos, Zhang2024IMsTech, avedillo2023operating}, probabilistic computing \cite{Deng2023PbitComp} and neural networks \cite{parihar2017vertex, velichko2019model, Velichko2017VO2NeuralNet, corti2021coupled, Nunez2021VO2neuralnet, maher2024highly, li2024computational, oh2021energy, shi2023integration, qiu2024reconfigurable}. These oscillators are characterized by simple design and nano-scale dimension\cite{qiu2024reconfigurable}, CMOS-compatibility \cite{maher2024highly} and robust synchronization properties \cite{shukla2014synchronized, velichko2018thermal, velichko2019investigation, velichko2021higher, pergament2016vanadium}. VO$_2$-based relaxation oscillators exhibit highly nonlinear dynamics \cite{shi2021dynamics}, driven by rapid phase transitions between the insulating and metallic states of VO$_2$ characterized by large hysteresis \cite{pattanayak2019tunable, Murtagh2020VO2Hysteresis} and memristive effects \cite{yi2018biological, wang2025nanoscale}, enabling adaptive nonlinear behaviors ideal for neuromorphic and unconventional computing applications \cite{Corti2020VO2NeuroComp, corti2021coupled, shamsi2021hardware, li2024computational}. 

However, despite promising capabilities, a relatively broad linewidth of VO$_2$-based relaxation oscillators on the order of tens of kHz, limits possible applications with VO$_2$-oscillators and their overall performance \cite{li2024computational, shukla2014synchronized}. Broad linewidth is especially detrimental for VO$_2$-oscillator-based Ising machines \cite{maher2024cmos, Zhang2024IMsTech, avedillo2023operating} since it disrupts injection locking, leading to the appearance of the phase slips \cite{litvinenko2021analog} which appear as random Ising spin flips. Neuromorphic computing systems \cite{Vodenicarevic2017NeurNetNoise, corti2021coupled} are also sensitive to broad linewidth in oscillatory neural networks because increased phase noise alternates collective synchronization patterns in the oscillator networks and reduces the signal-to-noise ratio required to detect the oscillator state. It also disturbs time-delay encoding in VO$_2$-oscillator-based spiking neural networks \cite{velichko2021higher, belyaev2019spiking, Yang2024NetworkVO2Osc, bohaichuk2020sizescal}. This degrades the ability of the VO2-based system to perform associative memory tasks, frequency-locked computing, and efficient phase-based information encoding.

In order to understand the mechanisms and corresponding noise types that contribute to the broad linewidth in VO$_2$-based oscillators, a comprehensive phase noise analysis has to be performed. Unlike conventional CMOS-based oscillators, VO$_2$-based relaxation oscillators rely on thermally induced phase transitions \cite{Maffezzoni2015VO2modelling, bohaichuk2020sizescal, brown2023electro} whose thresholds are susceptible to thermal fluctuations and demonstrate memristive behavior \cite{Naik2022VO2memristive, Shaobo2021StochasticVO2}, leading to increased jitter and phase noise \cite{Nunez2021VO2neuralnet}.

In this work, we study the phase noise composition of VO$_2$-based oscillators that leads to their broad spectral linewidth and study how the shape of external synchronization signals affects the phase noise of the oscillators. We find that external synchronization using a square-wave signal reduces susceptibility to thermal noise and improves the phase stability of VO$_2$-based oscillators, resulting in lower phase noise levels compared to a sinusoidal signal of the same amplitude. This improvement occurs because the sharp edges of the square wave steepen the approach to the switching threshold and shorten the stochastic incubation time.

\section{VO$_2$-based oscillator design and dynamics}

\begin{figure}[ht!]
    \centering
    \includegraphics[width=0.45\textwidth]{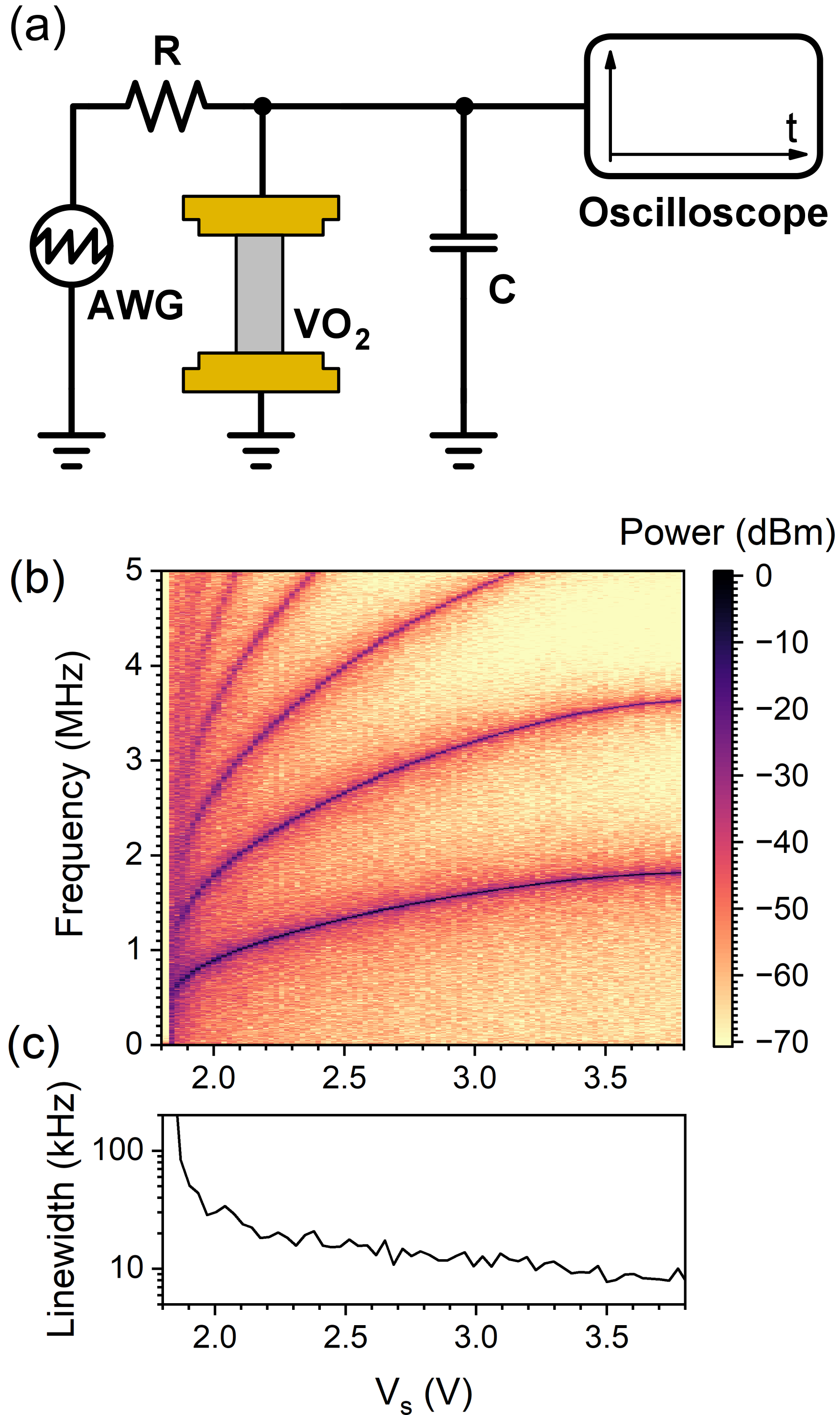}
    \caption{Basic characterization of the VO$_2$ relaxation oscillator. (a) Schematic of the oscillator circuit. (b) Spectrogram of the output signal as a function of the applied voltage, $V_{s}$. The lowest branch corresponds to the fundamental frequency described by Eq.~\ref{eq:Vout}. Higher-order harmonics appear at integer multiples of the fundamental frequency, exhibiting large amplitudes due to the strongly nonlinear shape of the time-domain signal. (c) Linewidth of the fundamental mode as a function of $V_{s}$.}
    \label{fig:figure1}
\end{figure}

Relaxation oscillators are typically implemented using CMOS-based nonlinear or switching elements such as inverters, comparators or Schmitt triggers \cite{geraedts2014towards}. In these designs, there is a capacitor that can be charged or discharged through a resistive element, and a comparator circuit is employed to monitor the voltage across the capacitor, switching between charging and discharging of the capacitor when specific voltage thresholds are crossed. Comparators in CMOS relaxation oscillators provide consistent state switching, allowing for stable period of oscillations and, therefore, relatively low jitter in a variety of electronic systems, including timing and clock generation circuits\cite{Denier2010}, analog-to-digital converters (ADC)\cite{Wang2000}, phase-locked loops (PLL)\cite{Fu2011}, and voltage-controlled oscillators (VCO)\cite{Hang2011, Oh2019}. The widespread use of CMOS elements for this class of oscillators is due to their integration capabilities, energy efficiency, and scalability. Moreover, designs that exploit CMOS technology offer the possibility to compensate for phase noise in comparators to achieve a practical figure of merit close to theoretically expected values \cite{geraedts2014towards}.

VO$_2$-based relaxation oscillators exploit the rapid insulator-to-metal transition (IMT) of vanadium dioxide as a non-linear switching element, effectively utilizing the IMT as a functional analog to a comparator \cite{shukla2014synchronized,parihar2015synchronization}. In this paper, we study the most common VO$_2$-based relaxation oscillator schematic, also referred to as the Pearson-Anson oscillator \cite{nandi2017temperature} (see Fig.1a). It consists of a simple RC-circuit needed for voltage hysteresis and a VO$_2$ film itself. When a voltage is applied, the capacitor C1 starts to charge while VO$_2$ remains in its insulating state until the voltage at VO$_2$ film crosses a critical threshold $V_{\text{MIT}}$, triggering a steep drop in resistance as the VO$_2$ material transitions to a metallic state \cite{maffezzoni2015modeling, shukla2014synchronized}. This abrupt change in resistance reduces the voltage drop on VO$_2$ in terms of direct current so that the capacitor C1 starts to rapidly discharge from $V_{IMT}$ through the VO$_2$ film. This discharge process is very short and, hence, has a minor contribution to the frequency and the observed phase noise in the system. Subsequently, when the voltage at the capacitor C1 drops below $V_{MIT}$, which cannot supply sufficient joule heating to maintain the metallic state, the VO$_2$ film relaxes backward from metallic to insulating state. After the moment of metal-to-insulator transition, the capacitor starts to charge till the level of the voltage drop on VO$_2$ in the insulator state:

\begin{equation}
    V_{\text{out}}(t) = V_{\text{st}} - (V_{\text{st}} - V_{\text{IMT}}) e^{- \frac{t}{\tau_{\text{ins}}}}
\label{eq:Vout}
\end{equation}

\begin{equation}
    V_{\text{st}} = \frac{R_{\text{ins}}}{R_{\text{ins}} + R} V_{\text{s}}
\end{equation}

\begin{equation}
    \tau_{\text{ins}} = \frac{R_{\text{ins}}}{R_{\text{ins}} + R} C
\end{equation}

where $C$ is the external capacitor, $R$ is the external resistor, $R_{\text{ins}}$ is the resistances of VO$_2$ in its insulating state, respectively, $V_{\text{s}}$ is the applied voltage, $V_{\text{st}}$ is the steady-state voltage,$V_{\text{IMT}}$ and $V_{\text{MIT}}$ are the threshold voltages for the IMT and MIT transitions, and $\tau_{\text{ins}}$ is the time constant of the exponential rise.

\begin{figure*}[ht!]
    \centering
    \includegraphics[width=\textwidth]{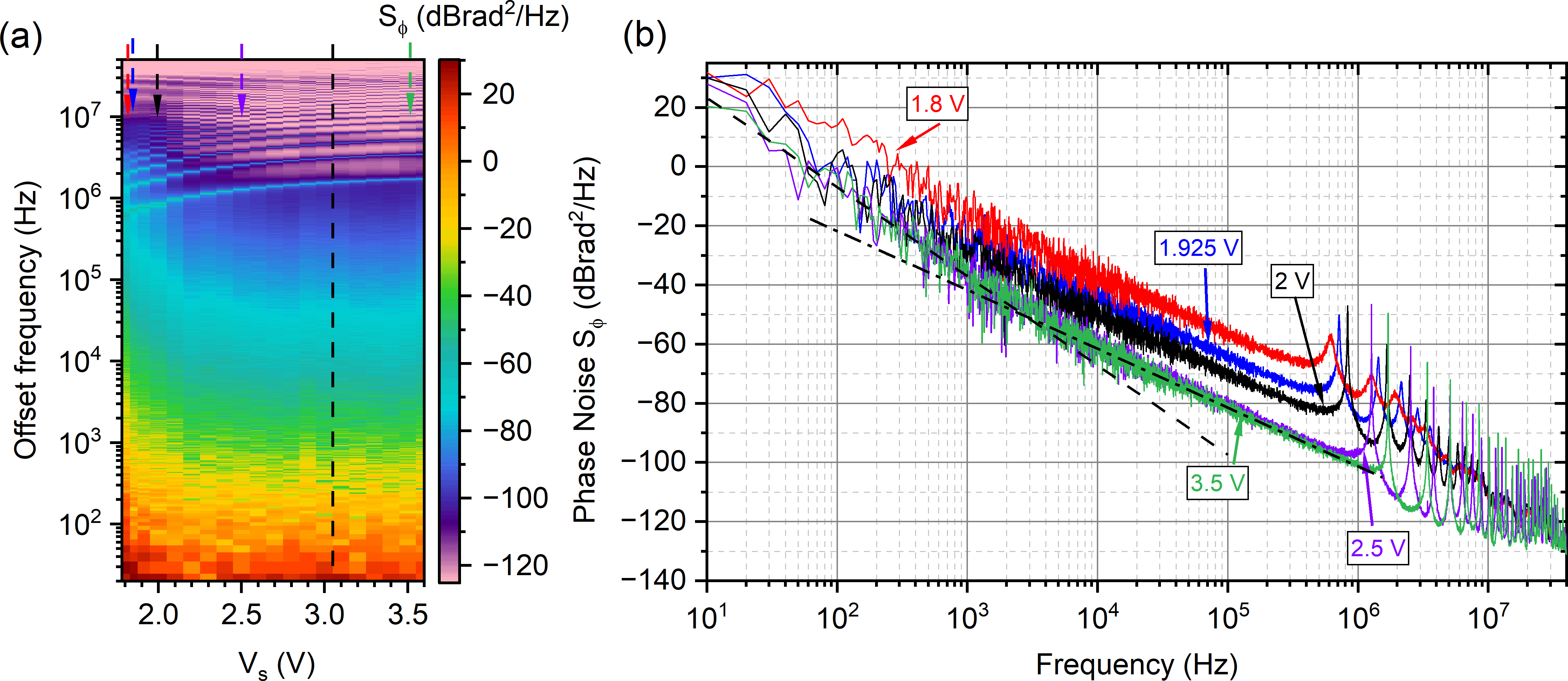}
    \caption{Phase noise characterization. (a) Phase noise spectrogram as a function of applied voltage $V_{s}$. (b) Phase noise power spectral density at different applied voltages. Note the increase in white noise components at lower voltages due to increased susceptibility to thermal fluctuations. At $V_{s}$ = 3.5 V and 2.5 V, the phase noise plot consists of two well pronounced regions, characterized by the slope $f^{-3}$ (highlighted with a dash line) corresponding to flicker frequency noise and by the slope $f^{-2}$ (highlighted with a dash-dot line) corresponding to white frequency noise. At $V_{s}$ = 1.8 V, the measured phase noise is dominated by the white frequency noise. The phase noise plots are extracted from 100-ms-long time traces digitized at 100 MS/s.}
    \label{fig:figure2}
\end{figure*}

Note that both $V_{\text{IMT}}$ and $V_{\text{MIT}}$ are strongly fluctuating, which causes frequency and phase noise in the oscillator. 

The frequency of oscillation is primarily controlled by the capacitor charging phase described in Eq.\ref{eq:Vout}, since the time required for the capacitor discharge through the VO$_2$ film in the metallic state is relatively short. Therefore, the overall oscillation frequency can be approximated as follows \cite{pattanayak2019tunable}:

\begin{equation}
    \frac{1}{f} \approx \tau_{\text{ins}} \ln \left( \frac{V_{\text{st}} - V_{MIT}}{V_{\text{st}} - V_{IMT}} \right)
\end{equation}

In Fig. 1b-c we demonstrate basic characterization of the VO$_2$-based oscillator. Fig.1b shows the spectrogram as a function of applied voltage $V_{\text{s}}$. The frequency dependence has an inverse logarithmic dependence according to Eq. 4. We can also see strong higher harmonics that originate from a strongly nonlinear relaxation process. The linewidth changes significantly with the applied voltage. In the next chapter, we investigate in detail why the linewidth and associated phase noise change with the applied voltage and how they can be improved.

\section{Phase Noise Analysis of Free-Running VO$_2$ Oscillator}

Figure 2 presents the phase noise characteristics of a free-running VO$_2$-based relaxation oscillator at various supply voltages. The left panel (Fig. 2 (a)) shows the phase noise $S_{\phi}$ in the form of a spectrogram as a function of offset frequency away from the carrier frequency and supply voltage $V_s$, while the right panel (Figure 2b) displays phase noise spectra at distinct supply voltages ranging from 1.8V to 3.5V. Let's first consider phase noise plot in Figure 2b at supply voltage $V_s = 3.5V$. At lower offset frequencies below $10^3$ Hz, the phase noise exhibits a $1/f^3$ slope typically classified as flicker frequency phase noise \cite{wittrock2020influence}. This noise is caused by low-frequency thermal and flicker noise in the VO$_2$ film, which results in frequency modulation of the oscillator. In the VO$_2$ relaxation oscillator, this flicker noise can originate from fluctuations in the threshold parameters of the insulator-to-metal transition. In the mid-range offset frequencies, approximately from $10^3$ Hz to $10^7$ Hz, the phase noise follows a $1/f^2$ behavior. The peaks above $5\times10^5$ Hz correspond to the main frequency and harmonics of the oscillator. Beyond $10^7$ Hz, the noise starts to flatten, transitioning to a white noise floor with a zero slope. This behavior is attributed to the thermal noise in the resistive elements of the circuit. In the VO$_2$-based oscillator, the white noise is primarily dominated by thermal noise contributions from the external resistor $R_1$ and the VO$_2$ device when it is in its insulator state.

\begin{figure*}[ht!]
    \centering
    \includegraphics[width=\textwidth]{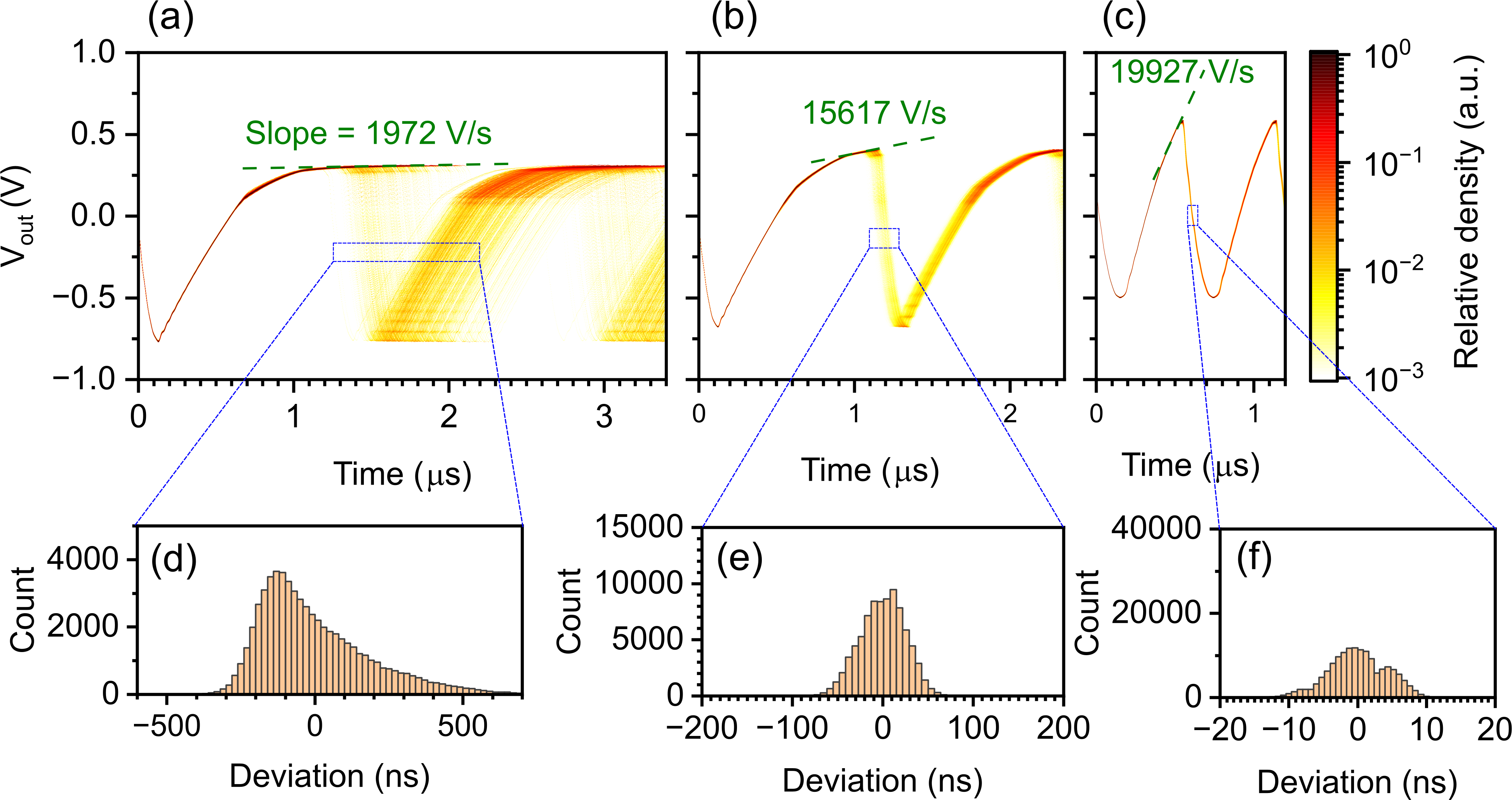}
    \caption{Time-domain traces and statistical jitter analysis of the $VO_{2}$ relaxation oscillator output at varying supply voltages: (a) $V_{s}$ = 1.81 V, (b) $V_{s}$ = 2.0 V, and (c) $V_{s}$ = 3.4 V. The top panels (a-c) show intensity-graded persistence maps where the color gradient represents the waveform probability density, visually highlighting cycle-to-cycle timing jitter. The DC component of the $V_{out}$ signal was removed (AC-coupled) to maximize the vertical resolution of the digitized signal. The bottom panels (d-f) show corresponding period length histograms. The results demonstrate that reducing the supply voltage decreases the slope of the voltage approach to the threshold, which significantly degrades switching stability and increases period variation.}
    \label{fig:figure3}
\end{figure*}

As the supply voltage decreases, the phase noise remains largely unchanged down to 2.2 V. Below this voltage, the white frequency noise ($1/f^2$ slope) begins to rise, gaining approximately 24 rad$^2$/Hz in phase noise power. Unexpectedly, the magnitude of the flicker frequency noise component ($1/f^3$ slope) remains constant; consequently, it is entirely masked by the elevated $1/f^2$ noise, which becomes the dominant contributor to the phase noise spectrum even at low offset frequencies down to 10 Hz.

\section{Time Traces and Jitter Analysis}
To better understand the impact of thermal noise on phase stability, we analyzed the time-domain traces of the oscillator output signal (Figure~\ref{fig:figure3}). To capture the small-amplitude voltage oscillations with higher vertical resolution, we configured the oscilloscope to block the constant direct current (DC) background voltage by using alternating current (AC) coupling. This approach provides increased sensitivity and reduces the noise from digitization since the DC offset of the signal is removed. The time traces are shown using an intensity-graded format. In this representation, the color gradient indicates the frequency of occurrence of the signal, providing a visual representation of the waveform probability density and highlighting the cycle-to-cycle timing jitter. Multiple periods are plotted on top of each other, starting from the same trigger level, forming a color map with a resolution of 1 mV and 10 ns. The relative density displays how frequently each point 1 mV by 10 ns was crossed with a time trace. In Figure~\ref{fig:figure3}d-e, we present the distribution of the period lengths in the form statistical histogram. We measured the jitter value by calculating the full-width half maximum (FWHM) of the distribution of period lengths at different supply voltages.

At supply voltage $V_{s} = 3.4$ V, the spread of time traces is almost negligible, and the jitter accounts only for 9.8 ns. When the supply voltage is reduced to 2.0 V, the spreading of the jitter intensity-graded trace becomes visible, which reflects in a significant increase in jitter of 50.7 ns. Then, at the supply voltage of 1.81 V we see significant blurring of the time trace with jitter degrading to 280.2 ns. Interestingly, the jitter histogram becomes asymmetric, indicating that the first-passage-time statistic governs the timing of the insulator-to-metal transition at low supply voltages. Because the supply voltage is close to the threshold, the system approaches the critical transition temperature significantly slower. In this shallow-drift regime, only rare high-amplitude thermal and voltage fluctuations can trigger the threshold crossing \cite{topalian2015resistance}, resulting in a characteristic right-skewed distribution with a pronounced tail towards longer switching times. The slopes of time traces before the insulator-to-metal transitions correlate well with both phase noise measurements and jitter.

\section{Synchronization with External Signal}
To mitigate the thermally induced phase fluctuations discussed above, we investigated synchronization of the VO$_2$ relaxation oscillator with an external signal at 2f. This choice is advantageous because it avoids direct spectral overlap between the intrinsic oscillation frequency and the external synchronization signal, while at the same time allowing for the two phase states separated by $\pi$ that are required for phase-binarized oscillator computing concepts such as Ising machines \cite{maher2024highly, avedillo2023operating, mcgoldrick2022SHNOIMs, houshang2022SHNOIM, gonzalez2024spintronic} and true number generators \cite{phan2024unbiased}. In the present device, this is particularly relevant because the spectral broadening observed when the oscillator operates autonomously originates primarily from the stochastic incubation interval preceding the insulator-to-metal transition (IMT), where the capacitor voltage approaches the switching threshold with a small slope and is therefore highly susceptible to thermal fluctuations. 

\begin{figure*}[ht!]
    \centering
    \includegraphics[width=\textwidth]{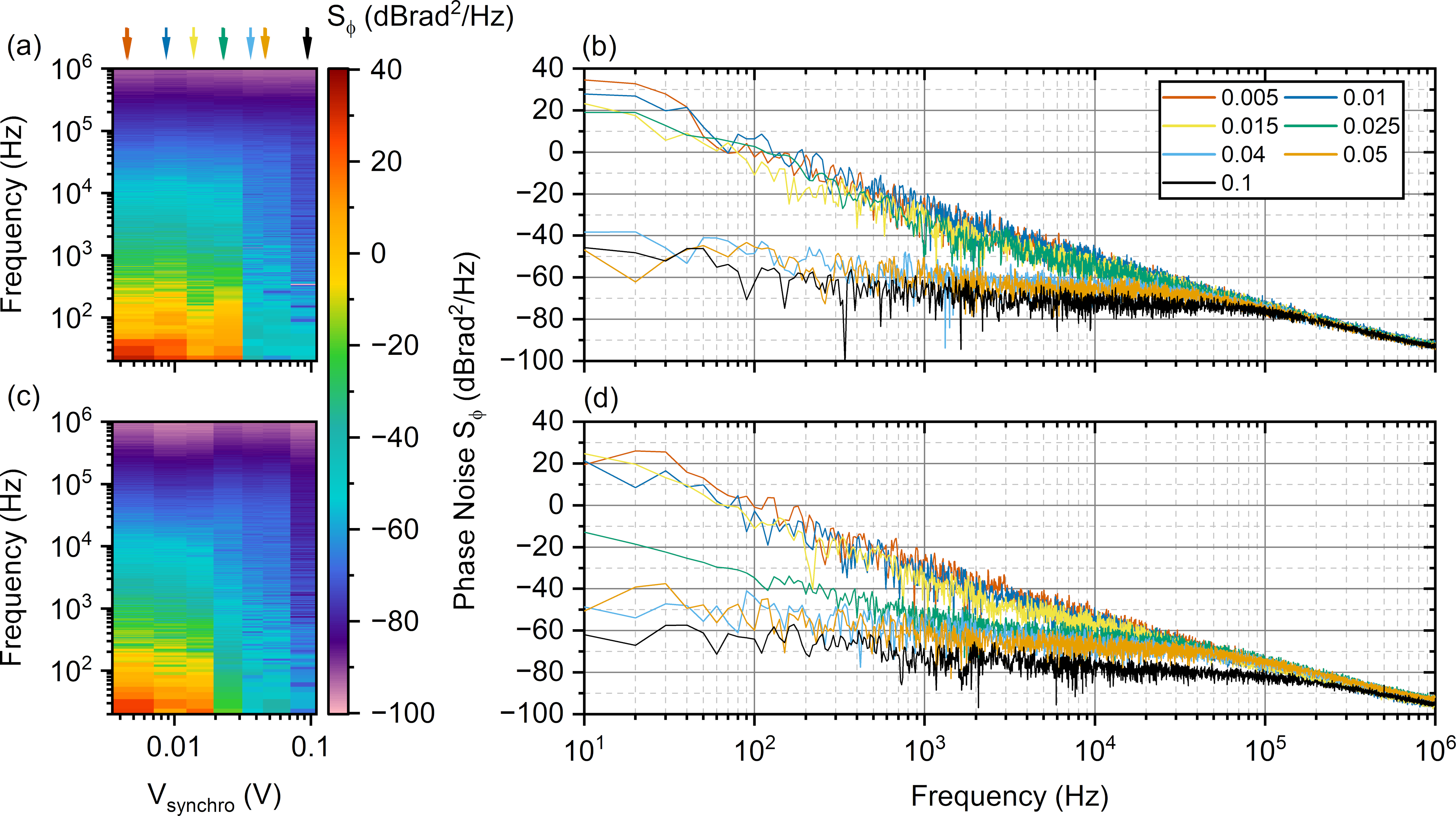}
    \caption{Phase noise suppression and synchronization of the $VO_{2}$ relaxation oscillator using external signals at 2f. (a, b) Phase noise spectrogram and phase noise spectral density plots for sinusoidal synchronization at various amplitudes. The colored arrows at the top of the spectrogram in panel (a) indicate the specific synchronization voltage levels ($V_{sync}$) that correspond to the identically colored phase noise spectral density curves plotted in panels (b) and (d). (c, d) Phase noise characteristics using a square-wave synchronization signal. The square-wave drive achieves partial injection locking and phase noise reduction at lower injected amplitudes than the sinusoidal drive. This improved stability occurs because the sharp edges of the square wave steepen the voltage approach to the switching threshold, thereby shortening the stochastic incubation time. Furthermore, the square-wave synchronization significantly increases the amplitude relaxation frequency, $\Gamma_{p}$, up to $1-2\times10^{5}$ Hz. In these plots, $\Gamma_{p}$ is visually identified as the corner frequency of the synchronization plateau in the fully locked regime, measured at the -3 dB drop from the low-frequency noise floor. This rightward shift of the corner frequency indicates a faster oscillator response to external signals.}
    \label{fig:figure4}
\end{figure*}

In Fig.~\ref{fig:figure4}, we show the evolution of the phase noise $S_{\phi}$ with the amplitude of the synchronization signal with sinusoidal and square-wave shape. Experimentally, the synchronization signals are formed by the same arbitrary waveform generator by adding a synchro signal $V_{sync}$ on top of the DC bias $V_s$ required for the sustainable auto-oscillations in the VO$_2$ relaxation oscillator. In both cases, increasing $V_{sync}$ first slowly shifts the phase-noise spectra downward by 10~$dBrad^2/Hz$ in the range of offset frequency $10^3-10^4$~Hz due to partial locking. At $V_{sync}$ of 0.04~V for sinusoidal shape, we see that the VO$_2$ relaxation oscillator gets fully locked, forming a synchronization plateau below the offset frequency $5\times10^5$~Hz, which defines the amplitude relaxation frequency $\Gamma_p$ of the oscillators. At a sinusoidal synchronization amplitude of $V_{sync} = 0.04$ V, the $VO_{2}$ relaxation oscillator achieves full injection locking. This is evidenced by the formation of a phase noise plateau at offset frequencies below $5\times10^{4}$ Hz, a boundary that corresponds to the amplitude relaxation frequency, $\Gamma_{p}$. In nonlinear oscillator dynamics, $\Gamma_{p}$ represents the rate at which the oscillator amplitude returns to its steady-state limit cycle following a perturbation. Practically, this parameter governs the transient response of the oscillator, defining how rapidly the device reacts to external injection signals by adjusting its amplitude and corresponding frequency \cite{litvinenko2021analog}. For coupled systems, such as oscillator-based Ising machines, $\Gamma_{p}$ is a critical performance metric because it directly dictates the convergence speed of the network and, consequently, the computational time-to-solution.

For sinusoidal synchronization, the reduction of phase noise is gradual at low drive amplitudes of 5–25~mV, and a pronounced transition to a phase-locking state with a phase-noise plateau occurs only when the injected amplitude reaches approximately 40~mV. The square-wave synchronization signal is more effective than the sinusoidal one for the same applied voltage amplitudes. In Fig.~\ref{fig:figure4}, a clear reduction of the low-offset phase noise is also visible at low $V_{sync}$ of around 5-15~mV. However, at 25~mV square-wave signal already leads to partial injection locking, showing a sudden drop of phase noise below $3\times10^3$~Hz. Interestingly, when the amplitude of a synchro signal $V_{sync}$ of square shape reaches 100~mV the amplitude relaxation frequency $\Gamma_p$ increases up to $1-2\times10^5$~Hz. This result is important for the design of the Ising machines since $\Gamma_p$ directly determines the speed of system convergence and, therefore, the time-to-solution. 

The superior performance of the square-wave drive can be understood directly from the switching physics of the VO$_2$ film. As shown by the jitter analysis, the timing uncertainty is governed by the ratio of the effective voltage noise to the slew rate at the switching threshold, $\delta t \sim \delta V/(dV/dt)_{th}$, and therefore increases strongly when the approach to the IMT becomes shallow. A square-wave edge rapidly increases the voltage across the VO$_2$ film and the corresponding Joule-heating power, thereby shortening the stochastic incubation time and steepening the trajectory toward the threshold. In contrast, a sinusoidal drive reaches the threshold more gradually, leaving the device longer in the thermally sensitive regime and allowing partial heat dissipation before the transition is completed. Because the free-running VO$_2$ oscillator is itself strongly non-sinusoidal and exhibits pronounced higher harmonics, the broader spectral content of the square-wave drive may also contribute to more efficient coupling to the nonlinear relaxation dynamics.

These results show that, for VO$_2$-based relaxation oscillators, the waveform of the synchronization signal is as important as its amplitude. Synchronization reduces the phase noise by introducing a deterministic phase-restoring action, but a sharp-edged injection-locking signal is considerably more efficient because it directly suppresses the stochastic threshold-crossing process. Moreover, square-wave synchronization leads to higher $\Gamma_p$ and faster oscillator response to external signals, making it especially attractive for Ising machines.

\section{Conclusions}
In this work, we performed a detailed experimental analysis of phase noise in VO$_2$-based relaxation oscillators and identified the physical origin of their linewidth broadening. We show that the dominant degradation mechanism at low oscillation frequencies is the stochastic incubation interval preceding the insulator-to-metal transition, where the capacitor voltage approaches the switching threshold with a reduced slope and is therefore highly susceptible to thermal fluctuations. As a result, the oscillator exhibits a strong increase in the white-frequency-noise contribution as the supply voltage approaches the transition threshold. We have confirmed it with time-domain measurements demonstrating that when the supply voltage is reduced, the temporal jitter increases dramatically and the period distribution evolves from narrow and nearly symmetric to broad and asymmetric, consistent with first-passage-time statistics governing the threshold crossing in the shallow-drift regime. The supplementary jitter analysis further confirms this picture by showing that the jitter follows the expected inverse dependence on threshold overdrive $\sigma \propto (V_s - V_{IMT})^{-1}$, yielding an extracted $V_{IMT} \approx 1.79$~V and a high-bias jitter floor of approximately 3.5 ns.

We further demonstrated that external synchronization at 2f provides an effective way to suppress these fluctuations and improve phase stability. Importantly, the results show that the waveform of the injected signal is a significant control parameter. Compared with sinusoidal injection, the square-wave synchronization signal reduces low-offset phase noise more efficiently, reaches the partially and fully locked regimes at lower amplitudes, and increases the amplitude relaxation frequency $\Gamma_p$ up to $1-2\times10^5$~Hz. This performance increase is explained by the sharp edges of the square wave, which steepen the approach to the IMT threshold, shorten the stochastic incubation time, and thereby suppress the threshold-crossing process responsible for the large linewidth in the free-running oscillator.

These findings provide both a physical understanding of phase noise in VO$_2$-based relaxation oscillators and a practical route for its control. More broadly, they show that the optimization of synchronization waveform and operating point is essential for the robustness, convergence speed, and phase stability of VO$_2$-oscillator networks intended for neuromorphic computing and oscillator-based Ising machines.

\section{Acknowledgment}
This work was supported by the Swedish Research Council (VR Grant No. 2024-01943) and by the Air Force Office of Scientific Research under award number \#FA9550-26-1B202. 

\appendix
\section{Device fabrication}
VO$_2$ films (about $100\,\mathrm{nm}$ thick) were grown on (012) Al$_2$O$_3$ substrates by reactive RF magnetron sputtering. The substrates were loaded into a high-vacuum chamber with base pressure $\sim 1 \times 10^{-7}\,\mathrm{Torr}$ and heated to $680\,^\circ\mathrm{C}$. During deposition, Ar flowed at $2.2\,\mathrm{sccm}$ and an O$_2$/Ar mixture (20\% O$_2$, 80\% Ar) flowed at $2.0\,\mathrm{sccm}$, giving a total pressure of $4\,\mathrm{mTorr}$. RF power was set to $100\,\mathrm{W}$ using a V$_2$O$_3$ target (discharge voltage $\sim 210\,\mathrm{V}$). The deposition lasted $30\,\mathrm{min}$ and produced $\sim 100\,\mathrm{nm}$ films. After growth, the samples were cooled to room temperature at $12\,^\circ\mathrm{C}/\mathrm{min}$ while keeping the same gas mixture.

The oscillator devices were fabricated by defining Cu/Pt pads using nanolithography and a lift-off process. First, the contact pad areas were patterned by electron-beam lithography using a Raith EBPG 5200 system and PMMA resist, with exposure doses of 290~\textmu C/cm$^2$. Subsequently, Cu (40~nm)/Pt (20~nm) thin films were deposited by magnetron sputtering. Finally, a lift-off process using 1165 remover was performed to obtain the final device structures. The  VO$_2$-film size was 500x100nm.

\section{Phase noise characterization}
The phase noise of the VO2-based oscillator in this study was measured via spectral analysis of the instantaneous phase signal obtained with a Hilbert transform method from time traces. This method \cite{litvinenko2021analog, litvinenko2023phase, bianchini2010}  allows to extract the instantaneous phase of a time trace signal by reconstructing a corresponding complex signal $x(t)$ from the measured signal $v(t)$, as follows:
\begin{equation}
    x(t) = v(t) + j \mathscr{HT} [v(t)] = A(t)e^{j\phi(t)}
\end{equation}
where $\mathscr{HT} [v(t)]$ is the Hilbert transform of $v(t)$, and $A_{out}(t)$ is time-varying amplitude of complex modulus $x(t)$ while $\phi(t)$ is its instantaneous phase.

The time domain signal $v(t)$ was captured with a 12-bit oscilloscope (Teledyne LeCroy WaveRunner 8208HD) with a sampling rate of 100 MS/s and $10^7$ number of points. Next, the Hilbert transform was done in MATLAB with the built-in function to obtain $x(t)$, and the instantaneous phase $\phi(t)$ signal was extracted with the "unwrap(angle())" command over the $x(t)$ signal. Then, a discrete Fourier transform of $\phi(t)$ was taken to visualize phase noise spectra $S_{\phi}$. 

The advantage of this method over conventional signal analyzers is the ability to characterize oscillators having an output signal with low signal-to-noise ratio and strong flicker noise. This method directly provides the power spectral density of phase fluctuations $S_{\phi}(f)$, Compared to the legacy $\mathcal{L}(f)$ metric typically measured by standard signal analyzers, $S_{\phi}(f)$  offers a more scientifically rigorous definition of phase noise \cite{rubiola2006measurement, rubiola2008phase}.

\section{Time domain and jitter analysis} 
Time domain signal from the oscillator output $v(t)$ was digitized using the same 12-bit oscilloscope (Teledyne LeCroy WaveRunner 8208HD). The oscilloscope channel was set to AC coupling to remove DC offset, increase the sensitivity of the measurements and reduce the influence of digitization noise. For jitter measurements we used $10^7$ points at a sampling rate of 100 MS/s to capture rapid changes of the insulator-to-metal and metal-to-insulator transitions.

Then, for each cycle, we calculated its temporal period using the same trigger level of the signal and compare it with the average period at the corresponding supply voltage $V_s$. The statistic is shown in the form of a statistical diagram. Detailed analysis of jitter characteristics at different supply voltages can be found in the Supplementary Materials.

\bibliography{references.bib}

\end{document}


\baselineskip24pt

\title{Supplementary Materials for} 
\subtitle{\textbf{Phase noise analysis and control of VO$_2$-based relaxation type oscillators}}

\maketitle 

\newpage

\section{Details of jitter analysis}

\begin{figure*}[ht!]
    \centering
    \includegraphics[width=0.7\textwidth]{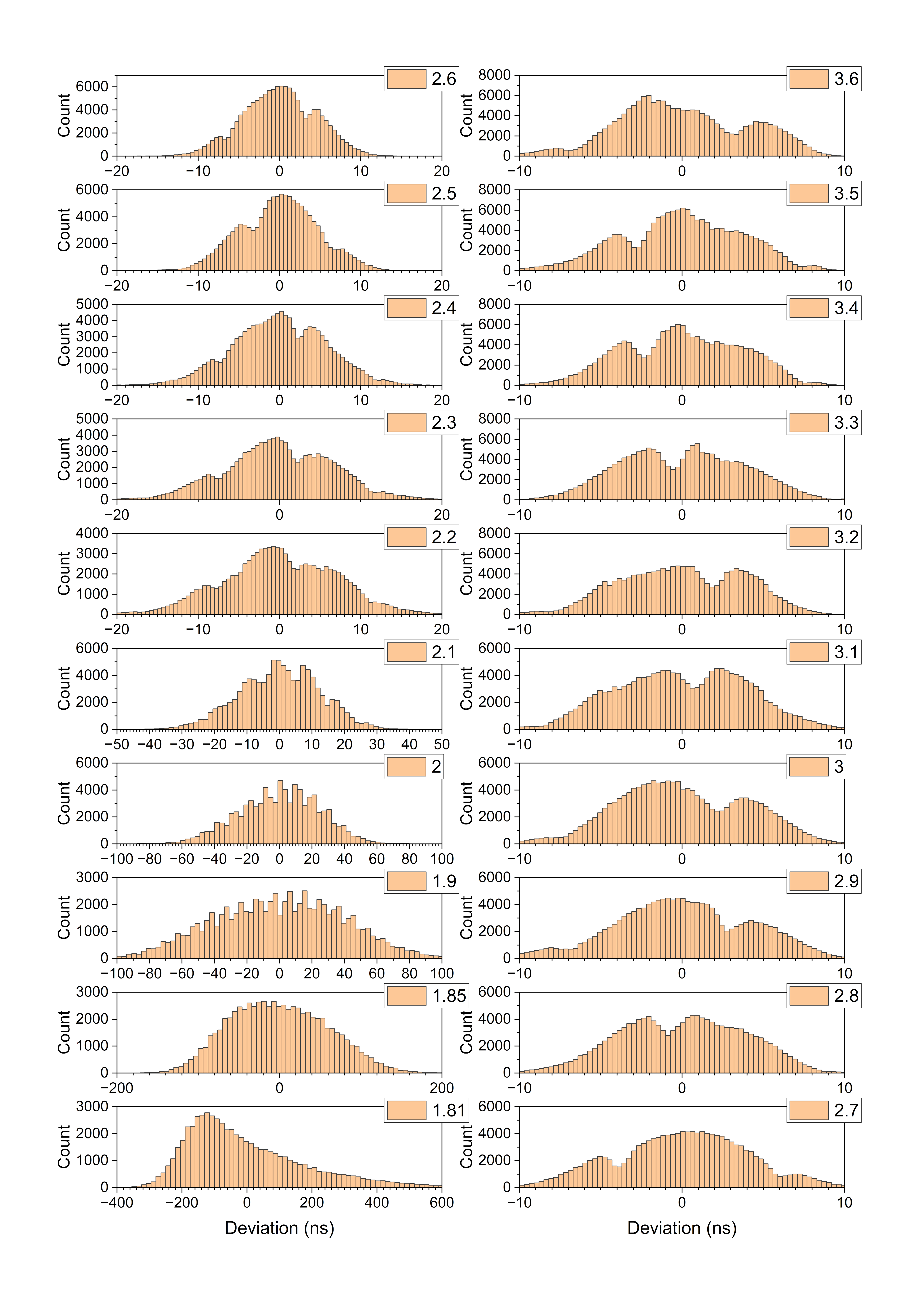}
    \caption{Detailed jitter analysis across varying supply voltages ($V_s$). The 20 panels illustrate the broadening and increasing asymmetry of the jitter distribution as the supply voltage is reduced, directly resulting from the reduction of the incubation voltage slope. The DC component of the $V_{out}$ signal was removed to utilize the full dynamic range of the oscilloscope. The distinct periodic ripples (10~ns spacing) visible on the histograms are an artifact of linearly interpolating the zero-crossing points from time traces sampled at a discrete rate of 100~MS/s.}
    \label{fig:Jitter}
\end{figure*}

To provide a comprehensive view of the switching stability, detailed jitter histograms are presented in Fig.~\ref{fig:Jitter}. The figure consists of 20 panels that map the evolution of the jitter across a wide range of the supply voltages $V_s$.

The experimental data demonstrate an expected dependence of the jitter distribution on the operating voltage. At higher supply voltages (e.g., $V_s > 3.0$~V), the oscillator is strongly driven, leading to a rapid crossing of the insulator-to-metal transition threshold. This steep voltage slope minimizes the system susceptibility to noise, resulting in narrow and highly symmetric jitter histograms. However, as the supply voltage is reduced towards the critical switching threshold (down to $\sim 1.8$~V), the incubation slope of the voltage trace decreases significantly. This slower approach to the transition point amplifies the impact of thermal and electrical fluctuations, causing the jitter distributions to broaden dramatically and develop an asymmetric shape. 

We also note that the periodic oscillations on the distribution curves of jitter histograms are associated with data processing artifact directly connected to how the jitter was computed. The raw time traces from the oscillator were digitized using an oscilloscope at a discrete sampling rate of 100~MS/s in order to capture long time traces, establishing a fixed temporal grid of exactly 10~ns between consecutive sample points. To precisely determine the oscillation period, the threshold crossing point was linearly interpolated using the slope of the sampled time trace. Because the raw data points are discretely spaced at 10~ns intervals, this interpolation heavily weights the probability density towards the sampled coordinates, resulting in the artificial periodic oscillations with an exact period of 10~ns visible on the jitter diagrams.

\begin{figure*}[ht!]
    \centering
    \includegraphics[width=0.5\textwidth]{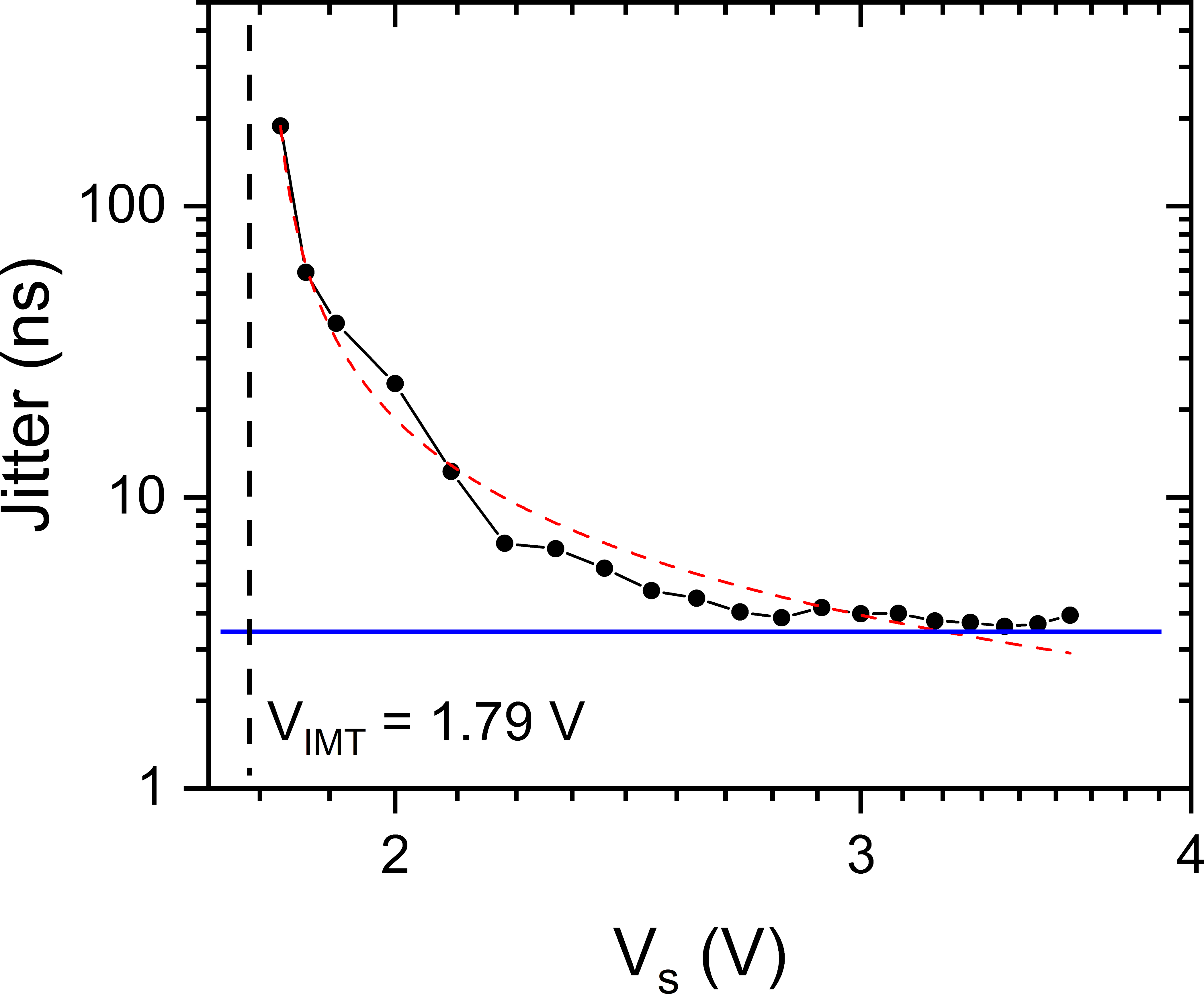}
    \caption{Dependence of Jitter on supply voltage $V_s$. As $V_s$ approaches the VO$_2$ insulator-to-metal transition threshold $V_{IMT}$, the timing jitter increases dramatically, following an inverse relationship: $\sigma \propto (V_s - V_{IMT})^{-1}$. The red dash line shows a fit to this inverse model, yielding an extracted threshold of $V_{IMT} = 1.79$ V. At supply voltages above $\sim 3$ V, the jitter asymptotes to approximately 3.5 ns.}
    \label{fig:Jitter_vs_Vs}
\end{figure*}

We also plotted in Fig.~\ref{fig:Jitter_vs_Vs} and analyzed absolute timing jitter as a function of the supply voltage, $V_s$. In a relaxation oscillator, the temporal jitter is defined by the ratio of the noise voltage to the voltage slew rate ($dV/dt$) precisely at the switching threshold. Because the incubation slew rate is proportional to the overvoltage ($V_s - V_{IMT}$), the timing jitter is theoretically expected to scale inversely with the threshold overdrive: $\sigma \propto (V_s - V_{IMT})^{-1}$. Fitting our experimental data to this inverse-linear model yields an excellent agreement with an extracted critical threshold voltage of $V_{IMT} \approx 1.79$~V. This fit confirms that the drastic amplification of phase noise observed at lower supply voltages is primarily driven by the flattening of the incubation voltage slope, which severely heightens the system's susceptibility to intrinsic voltage fluctuations during the critical nucleation phase. Moreover, we can see from the plot that the jitter flattens out at the jitter level of 3.5 ns. 

\newpage